\documentclass[12pt,preprint]{aastex}

\usepackage{graphicx} 

\shorttitle{QPOs in H1743--322}
\shortauthors{Jeroen Homan et al.}

\def\1746{H1743--332}
\newcommand{\numax}{$\nu_{max}$}

\begin{document}

\title{High- and low-frequency quasi-periodic oscillations in the X-ray light curves of the black hole transient \1746}

\author{Jeroen Homan\altaffilmark{1}, Jon M.~Miller\altaffilmark{2,5}, Rudy Wijnands\altaffilmark{3}, Michiel van der Klis\altaffilmark{3}, Tomaso Belloni\altaffilmark{4}, Danny Steeghs\altaffilmark{2} and Walter H.G.~ Lewin\altaffilmark{1}}

\altaffiltext{1}{MIT Center for Space Research, 70 Vassar Street, Cambridge, MA 02139}
\altaffiltext{2}{Harvard-Smithsonian Center for Astrophysics, 60 Garden Street, MA 02138}
\altaffiltext{3}{Astronomical Institute "Anton Pannekoek", University of Amsterdam, Kruislaan 403, 1098 SJ Amsterdam, The Netherlands}
\altaffiltext{4}{INAF/Osservatorio Astronomico di Brera, Via E.~Bianchi 46, 23807 Merate (LC), Italy }
\altaffiltext{5}{NSF Astronomy \& Astrophysics Fellow}

\begin{abstract}

We present a variability study of the black hole candidate and X-ray
transient \1746 during its 2003/2004 outburst. We analyzed five Rossi
X-ray Timing Explorer observations that were performed as part of a
multi-wavelength campaign, as well as six observations from the early
rise of the outburst. The source was observed in several black hole
states and showed various types of X-ray variability, including high
frequency quasi-periodic oscillations (QPOs) at 240 Hz and 160 Hz
(i.e.\ with a 3:2 frequency ratio), several types of low frequency
QPOs, and strong variations on a time scale of a few hundred seconds.
The discovery of high frequency QPOs in \1746 supports predictions
that these QPOs should be more easily observed in high inclination
systems. In one of our observations we observed a transition in count
rate and color, during which we were able to follow the smooth
evolution of the low-frequency QPOs from type-B to type-A. We
classify the X-ray observations and QPOs, and briefly discuss the 
QPOs in terms of recently proposed models.

\end{abstract}

\keywords{stars: individual (\1746) --- stars: oscillations ---X-rays: binaries --- X-rays: stars}

\section{Introduction}

Outbursts of transient X-ray binaries provide excellent opportunities
to study the properties of accretion flows onto compact objects over
a large range of mass accretion rates. Black hole transients are in
general more luminous and show more pronounced changes in their
spectral and variability properties than neutron star transients.
Past observations of black hole transients have revealed an
intriguing variety of correlated spectral and variability properties
that is often described in terms of so-called black holes states and
transitions between them \citep{tale1995,va1995a,mcre2003}. Three
such states are generally recognized: the hard state, the soft (or
thermal dominant) state, and the very high (or steep power law) state
\citep[see][for a recent discussion]{mcre2003}. It has been found
that not only the X-ray properties change from one state to the
other, but that major changes in the radio
\citep[e.g.][]{fecotz1999}, optical \citep[e.g.][]{jabaor2001b} and
IR \citep[e.g.][]{hobuma2004} occur as well. As these different
wavelengths probe different regions and mechanisms in the accretion
flow it is important to obtain simultaneous data in order to
construct a consistent picture of the accretion flow. To this end we
organized a multi-wavelength campaign -- using RXTE, {\it Chandra},
radio  and optical observations -- to study a black hole transient in
outburst. This program was triggered by an outburst of \1746 in March
2003. Results of the analysis of the X-ray spectra are reported by
\citet{miraho2004}. The results of radio observations made in partial
support of this program will be reported in a separate paper by Rupen
et al. (2004, in prep.), and results from optical and IR observations
will be reported in Steeghs et al. (2004, in prep.).  Here we report
on the variability properties of the RXTE observations, focusing on
the quasi-periodic oscillations (QPOs).  

Different types of black hole QPOs were already found with {\it
Ginga} in several sources \citep[e.g.][]{mikiki1991,tadomi1997}, but
it was not until a few years ago that they were classified in a
systematic way. \citet{wihova1999} and \citet{resomu2002} recognized
three types of low frequency ($<10$ Hz) QPOs in XTE J1550--564, which
they named type-A, B and C. The strongest of these, type-C, have
frequencies ranging from tens of mHz to $\sim$10 Hz. They are always
accompanied by strong band limited noise (i.e. with a clear break in
the power as function of frequency) and are associated with the hard
state and transitions to and from the soft state. Type-A and B QPOs,
only found in the very high/steep power law state, are confined to a
narrow frequency range of $\sim$4--9 Hz and are accompanied by weak
power law noise (red noise). Type-B QPOs are more coherent and have a
higher harmonic content. All three QPO types also have unique time
lag properties \citep{wihova1999,resomu2002}, showing both hard and
soft lags, that are hard to explain with simple Comptonization
models. Unfortunately, the large variety of black hole QPOs is not
well understood. Almost every black hole X-ray transient has shown
one or more of these three QPO types, although not all observed QPOs
can be classified within this scheme \citep[see e.g.][]{resomu2002}.
For example, a weak QPO around 20 Hz has been observed during the
soft states of XTE J1550--564 \citep{howiva2001} and GRO J1655--40
\citep{somcre2000a}. Other types of low frequency QPOs were found in
4U 1630--47 \citep[$\sim$0.1 Hz;][]{dibe2000} and GRS 1915+105
\citep[$\sim$0.01 Hz;][]{moregr1997}. These QPOs were often present
simultaneously with type-A, B or C QPOs and possibly represent fast
transitions between different (sub-)states \citep[see e.g. the
dynamical power spectra in][]{mumore1999}. High frequency ($>50$ Hz)
QPOs have also been found in a small number of systems. They appear
to be stable in frequency and in a few cases two harmonically related
peaks (with frequency ratios of 3:2 or 5:3) have been observed at the
same time \citep{st2001a,st2001b,miwiho2001} which have been used to
constrain the spin of the black hole. 

\1746 was discovered in August 1977 with Ariel 5
\citep{kaho1977a,kaho1977b} as a new transient ``with a spectrum
somewhat harder than the Crab" \citep[HEAO-1
observations]{dobrfa1977}. The outburst lasted until at least March
1978 \citep{woshjo1978}. \citet{whma1983} identified the source as a
potential black hole candidate based on its ultra-soft X-ray
spectrum, which was similar to that of other black hole candidates in
their soft state.  High energy observations during the 1977/78
outburst also revealed the presence of a high energy tail in the
spectrum \citep{colela1984}. A new outburst of the source was
detected\footnote{The new detection led to the addition of the
following source identifiers: IGR J17464--3213 and XTE J1746(4)--322.
In the literature the source is sometimes also referred to as
H1741--32. In the RXTE archive the source can be found as XTE
J1746--319} with Integral \citep{rechca2003} and RXTE
\citep{masw2003a,masw2003b} in March 2003, that lasted until early
2004 (see Figure \ref{fig:asm}) . The only reported detections of the
source between the 1977/78 and 2003/04 outbursts were with EXOSAT in
1984 \citep{repaha1999} and the TTM/COMIS  telescope (2--30 keV)
on-board Mir-Kvant in 1996 \citep{emalsu2000}. Soon after the
Integral and RXTE detections, radio \citep{rumidh2003}, infrared
\citep{banaiw2003}, and  optical \citep{stmika2003} counterparts were
found. Integral and RXTE observations suggest that the source went
through several states during its outburst
\citep{masw2003a,homiwi2003,krchca2003,grlusu2003,toka2003,pakuoo2003}.
 \citet{masw2003a} found a 0.05 Hz QPO with an rms amplitude of 25\%
during the early stages of the outburst, typical of the hard state.
\citet{homiwi2003} already reported on QPOs around 6 Hz \citep[see
also][]{spstka2004} and the discovery of a QPO at 240 Hz. QPOs around
8 Hz were found by \citet{toka2003} when the source was making a
transition back to the hard state. In this paper we present a more
detailed study of the high frequency QPO and the evolution of the
lower frequency QPOs during a rapid transition in the light curve.

\section{Observations and data analysis}\label{sec:obs}

We observed \1746\ with RXTE \citep{brrosw1993} on five occasions
between 2003 May 1 and 2003 July 31. These observations,
which were all part of program P80135, spanned multiple RXTE observation
IDs and most of these  contained multiple RXTE orbits. A log of the
observations can be found in Table \ref{tab:log}. Observations 1--3
and 5 were performed simultaneously with {\it Chandra} observations.
Spectral results from these combined observations are reported by
\citet{miraho2004}. The variability study in this paper is solely
based on data from the proportional counter array 
\citep[PCA, consisting of five similar units;][]{jaswgi1996} on-board RXTE. The PCA data we used were in
the following modes. {\tt Standard 2}, which has a time
resolution of 16 s and covers the 2--60 keV PCA effective energy
range with 129 energy channels. {\tt SB\_125us\_8\_13\_1s} and {\tt
SB\_125us\_14\_35\_1s}, which have a time resolution of $2^{-13}$ s
($\sim$122 $\mu s$) and cover the 3.3--5.8 keV and 5.8--14.9 keV
bands with one channel. And finally, {\tt E\_16us\_16B\_36\_1s}, which
has a time resolution of $2^{-16}$ s ($\sim$15 $\mu s$) and covers
the 14.9--60 keV band with 16 channels. {\tt Standard 2} data from
proportional counter unit (PCU) 2 were background subtracted and then
used to create light curves and color curves, with color defined as
the ratio of counts in the  9.5--17.9 keV and 2.5--5.8 keV bands
({\tt Standard 2} channels [ranging from 1 to 129] 20--40 and 3--10). 

Fast Fourier transforms (FFTs) of the high time resolution data from
all the active PCUs\footnote{For reason of detector preservation some PCUs
were switched off}  were performed to create power spectra with a
frequency range of $0.016-4096$ Hz in the 5.8--20.9 keV band
(absolute channels [ranging from 0 to 254] 14--49). This energy range
was chosen since high-frequency variability in black hole low-mass
X-ray binaries is in general most significantly detected in similar
energy ranges. The data were not background-subtracted, and no dead
time corrections were applied prior to the FFTs; the effects of dead
time were accounted for by our power spectral fit function (see
below). Power spectra were averaged based on time, colors, or count
rate, and normalized according to the recipe described in
\citet{va1995b}. 

As in our work on XTE J1650--500 \citep{hoklro2003} we fit (part of)
the resulting power spectra with a sum of Lorentzians, each given by
$P(\nu)=(r\Delta/\pi)[\Delta^2 + (\nu-\nu_0)^2]^{-1}$, where $\nu_0$
is the centroid frequency, $\Delta$ the half-width-at-half-maximum,
and $r$ the integrated fractional rms (from $-\infty$ to $\infty$).
Instead of $\nu_0$ and $\Delta$ we will quote the frequency at which
the Lorentzian attains its maximum in $\nu P(\nu)$, $\nu_{max}$, and
the quality factor, Q, where $\nu_{max}=\nu_0 (1 + 1/4Q^2)^{1/2}$ and
$Q=\nu_0/2\Delta$ \citep{bepsva2002}. The fractional rms amplitudes
quoted in this paper are the integrated power between 0 and $\infty$.
The dead time-modified Poisson noise was approximated by a constant.
Errors on fit parameters were determined using $\Delta\chi^2=1$.
Upper limits on the strength of the Lorentzians were determined  by
fixing the Q and/or \numax, but not the rms amplitude, to  values
similar to those obtained in another power spectrum and using
$\Delta\chi^2=2.71$ (95\% confidence). All quoted significances for
QPOs are for single trial. When fitting high frequency QPOs, we only
fitted above $\sim$25 Hz, approximating the noise below 100 Hz with a
simple power law. For observations with strong low frequency QPOs we
also performed a study of the phase lags between the 2--5.8 keV
(soft) and 5.8--14.9 keV (hard) bands, using methods described in
\citet{wihova1999}.

\section{Results}

\subsection{Light curves and colors}

\begin{figure} \centerline{\includegraphics[width=12cm]{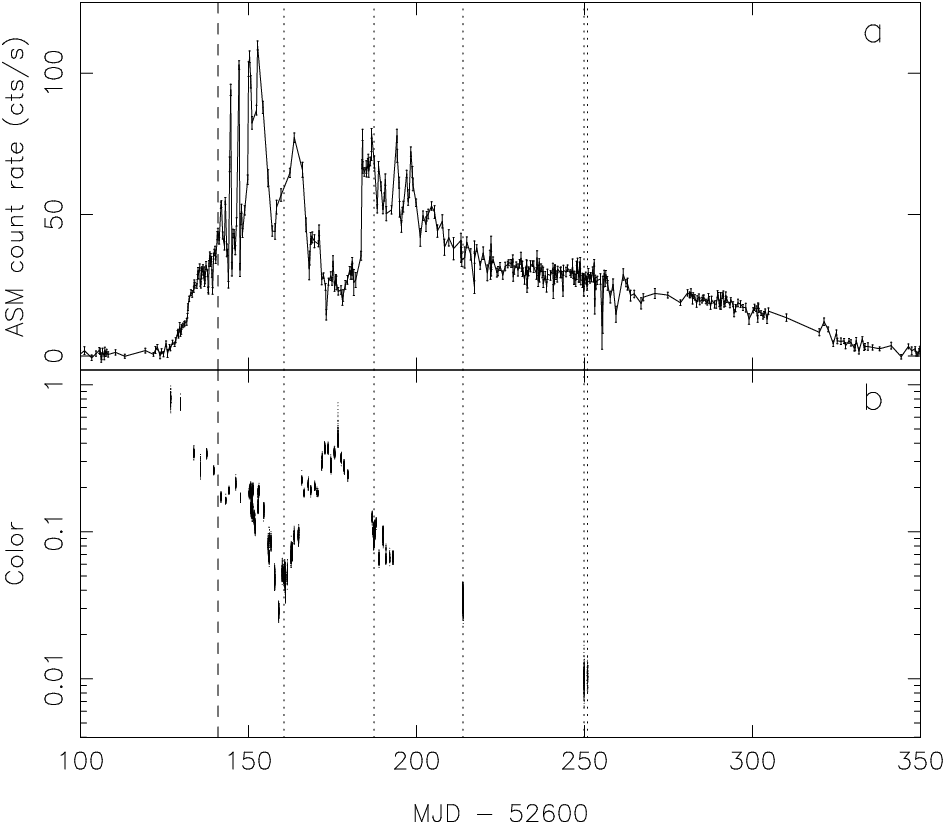}}
\caption{Light and color curves of the 2003/2004 outburst of
H1743--332. Panel (a) shows the RXTE/ASM light curve
($\sim$2--12 keV) and panel (b) shows the color curve, obtained from
RXTE/PCA data (higher values indicating a harder spectrum). The five
dotted lines mark the times of our (quasi-)simultaneous RXTE/{\it
Chandra} observations. Observations a--f were all made before the
dashed line.}\label{fig:asm} \end{figure}

Figure \ref{fig:asm} shows the RXTE all-sky monitor light curve
($\sim$2--12 keV) of the 2003/2004 outburst of \1746\ together with
X-ray colors (as defined in the Section \ref{sec:obs}) from PCA data.
The outburst is not of the 'classical' fast-rise-exponential-decay
type, showing complex behavior similar to that observed in XTE
J1550--564 and XTE J1859+226. The dates of our observations are
marked by the vertical dotted lines. In panel \ref{fig:asm}b, which
shows the PCA colors, we have also included data from  other
observations of \1746\ that were publicly available at the time of
writing. The six observations before the vertical dashed line (from
program P80138,  hereafter observations a--f) were also part of our
variability analysis (see Section  \ref{sec:res_low}). These
observations clearly show that the source was much harder during the
early phase of the outburst than in our observations. Variability
studies of the other public observations can be found in
\citet{spstka2004} and Remillard et al. (2004, in prep.)  Average
PCA count rates in the 2--60 keV band and the strength of the
broad-band variability are given in Table \ref{tab:log} for each of
our five observations. 

\begin{figure}
\centerline{\includegraphics[width=12cm]{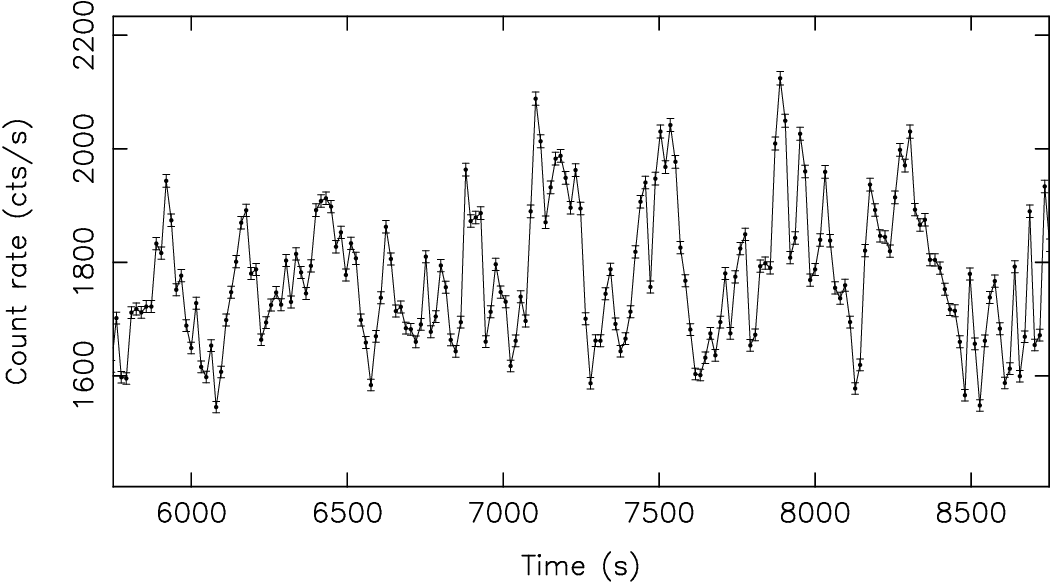}}

\caption{An example of strong variability on a time scale of a few hundred seconds as was observed in observations 1--3. This light curve (2--60 keV) is from observation 1. It is background subtracted and has a time resolution of 32 s.}\label{fig:flaring}
\end{figure}

\begin{figure}
\centerline{\includegraphics[width=12cm]{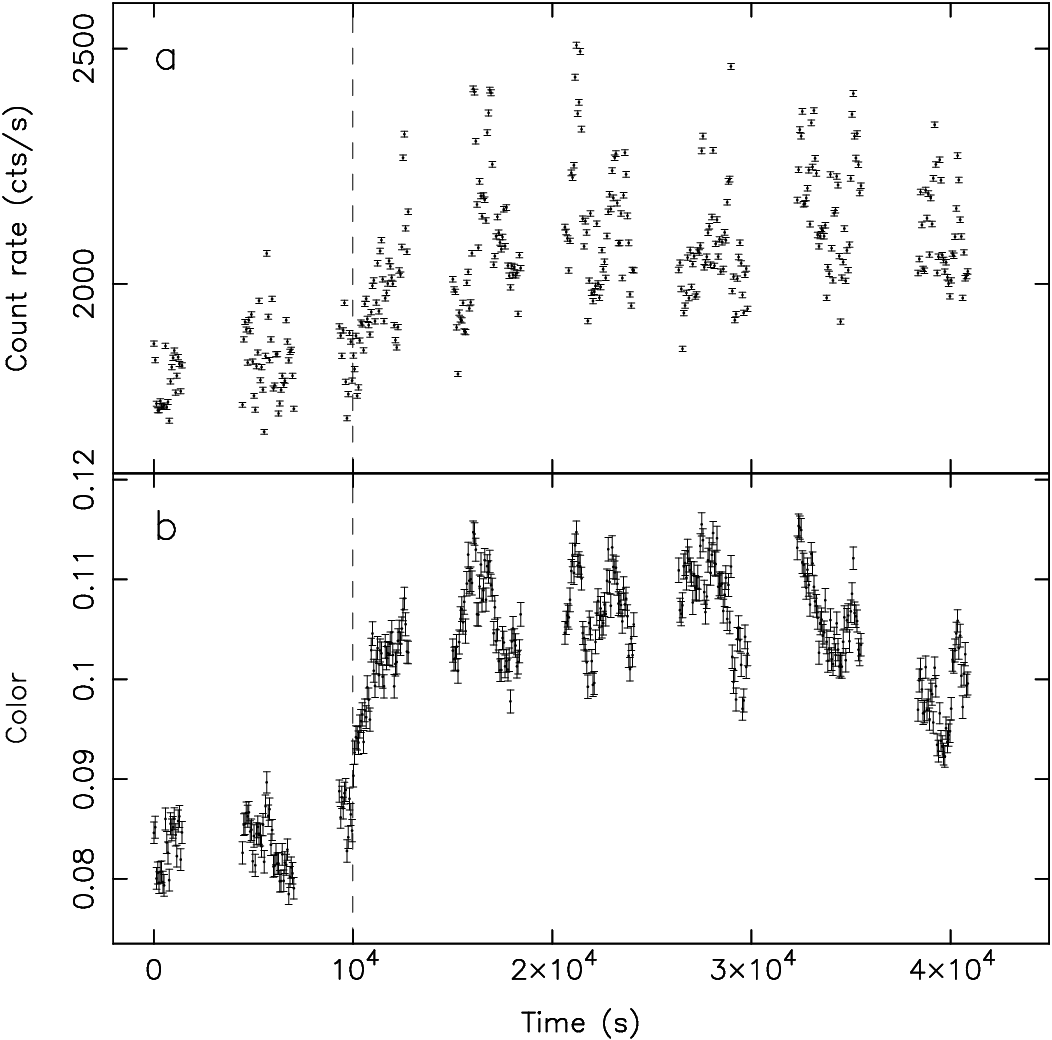}}

\caption{Light and color curves of the transition that was seen in observation 2. Panel (a) shows the 2--60 keV light curve and panel (b) the color curve. Both curves are background subtracted and have time resolutions of 64 s. The dashed line indicates the start of the transition.}\label{fig:curves}
\end{figure}

The PCA light curves of observations 1--3 showed strong variability
with a typical time scale of a few 100 seconds -- an example from
observation 1 can be seen in Figure \ref{fig:flaring}. This
variability was also observed at lower energies with {\it Chandra},
but it is stronger at energies above 10 keV, which can be probed by
RXTE \citep[see Figs.~2 and 5 in ][]{miraho2004}. Observation 2 was
characterized by a transition in count rate and color that occurred
$\sim$10 ks after the start of the observation (see
Fig.~\ref{fig:curves}). A sudden jump in
color, indicated by the dashed vertical line, occurred within 100 s,
but the overall change took about 1000 s. The transition in count
rate is less pronounced, but the difference in average count rate
before and after is considerable (see Table \ref{tab:log}). At the
end of the observation the source seemed to become softer again. The
light curves of observations 4 and 5 were relatively featureless,
showing neither the strong few hundred second variations nor sudden
transitions.

\subsection{Low frequency QPOs} \label{sec:res_low}

\begin{figure}
\begin{center}
\includegraphics[height=5.5cm]{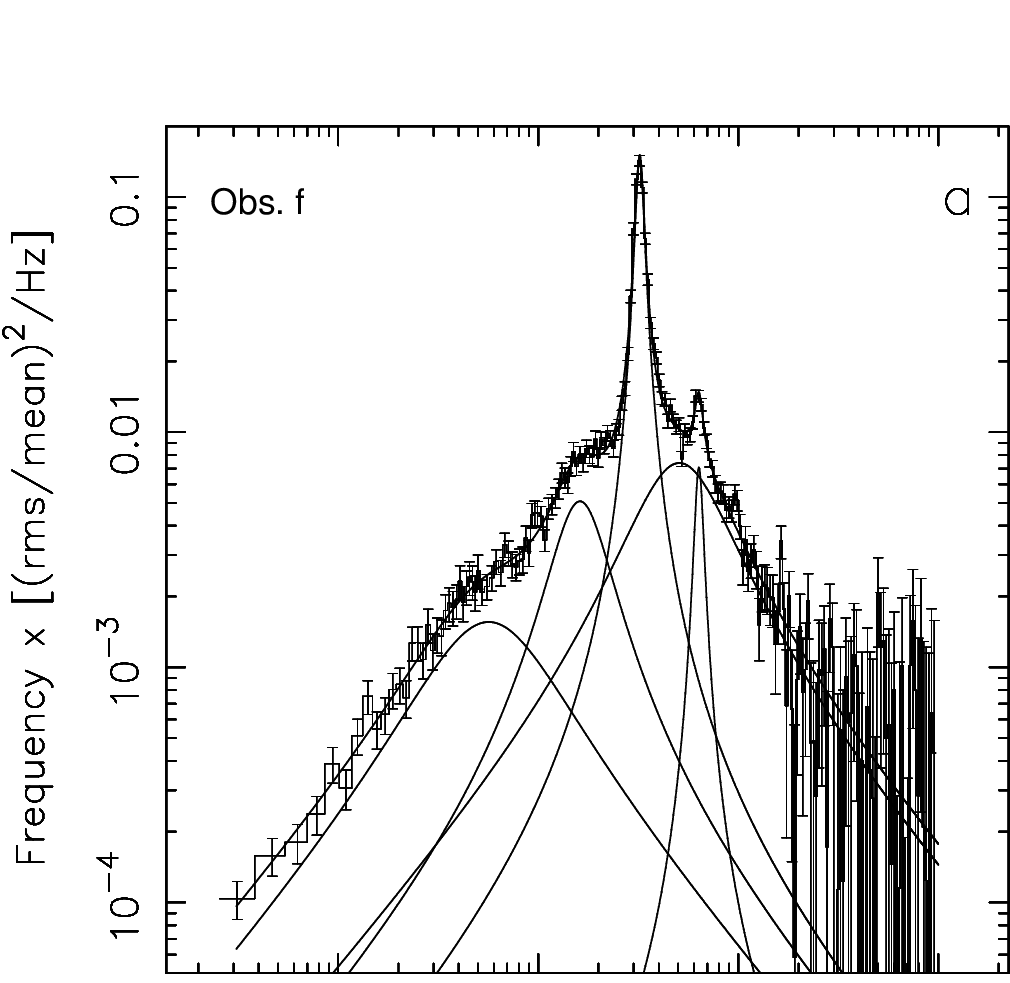}
\includegraphics[height=5.5cm]{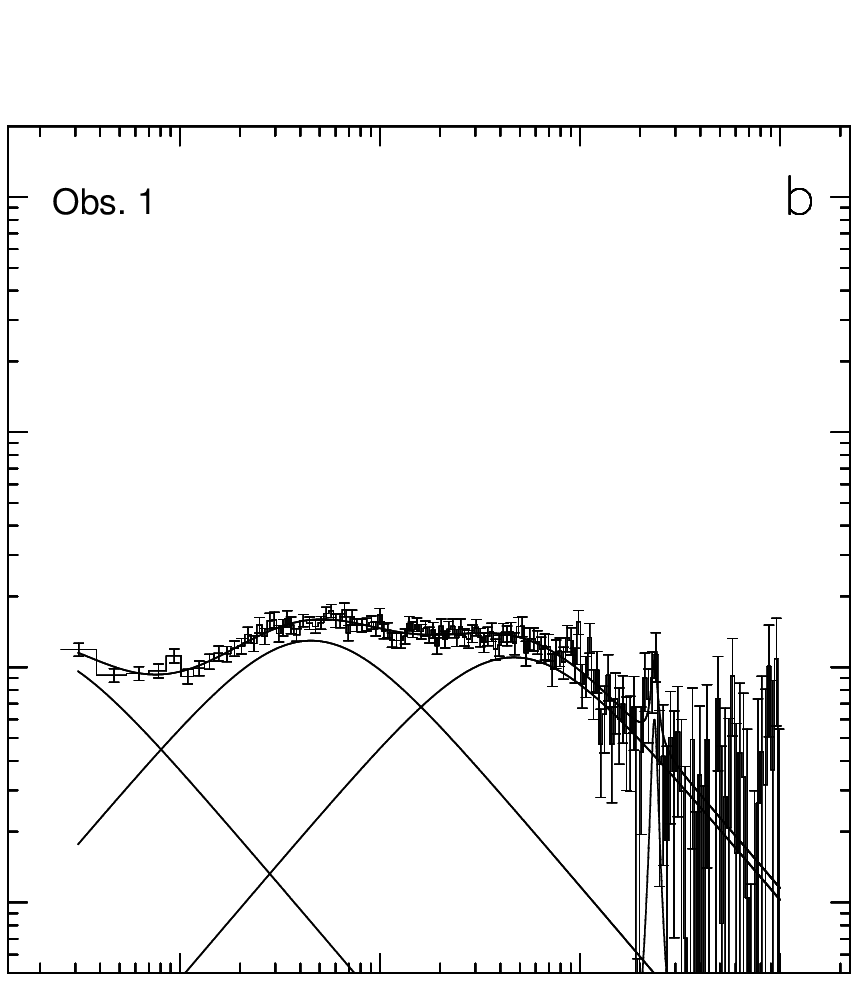}
\includegraphics[height=5.5cm]{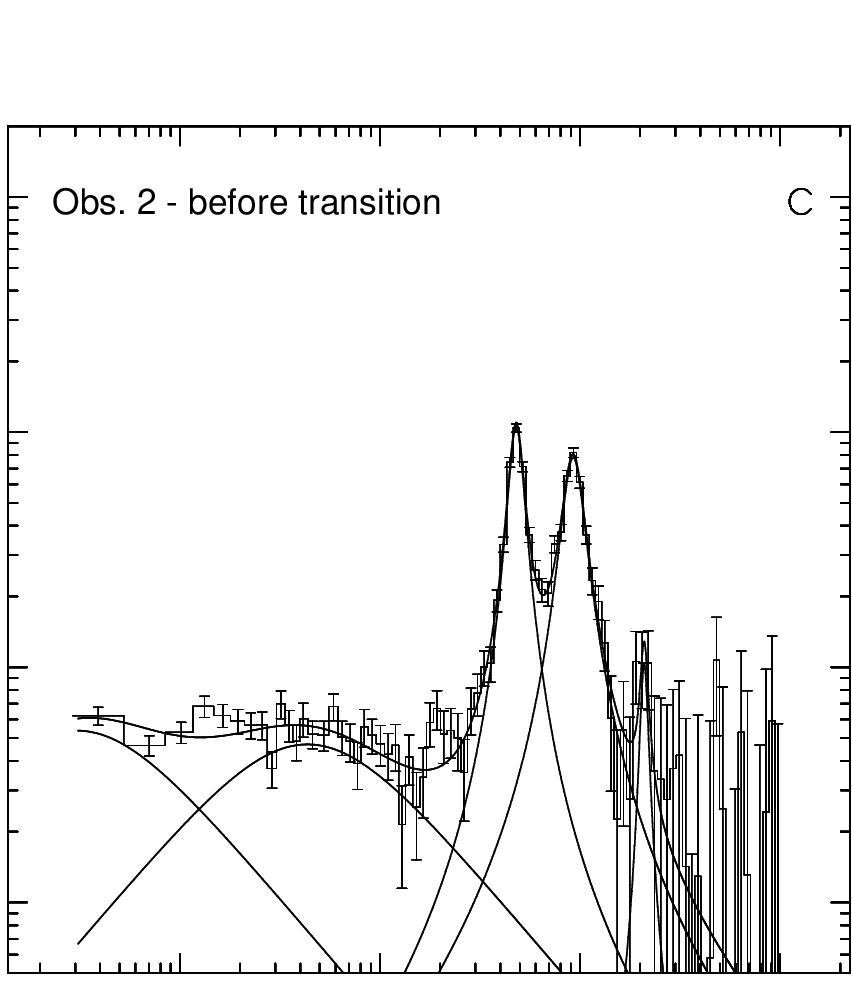}\\
\includegraphics[height=5.5cm]{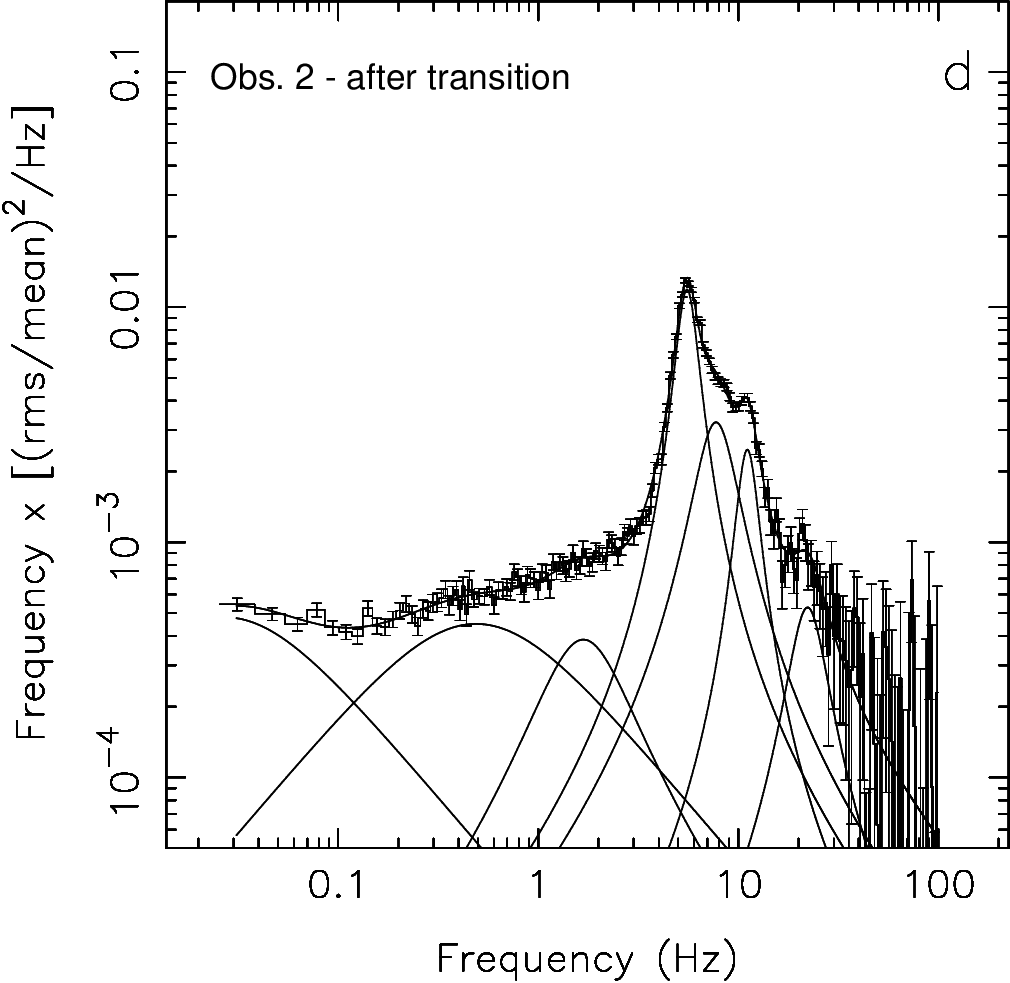}
\includegraphics[height=5.5cm]{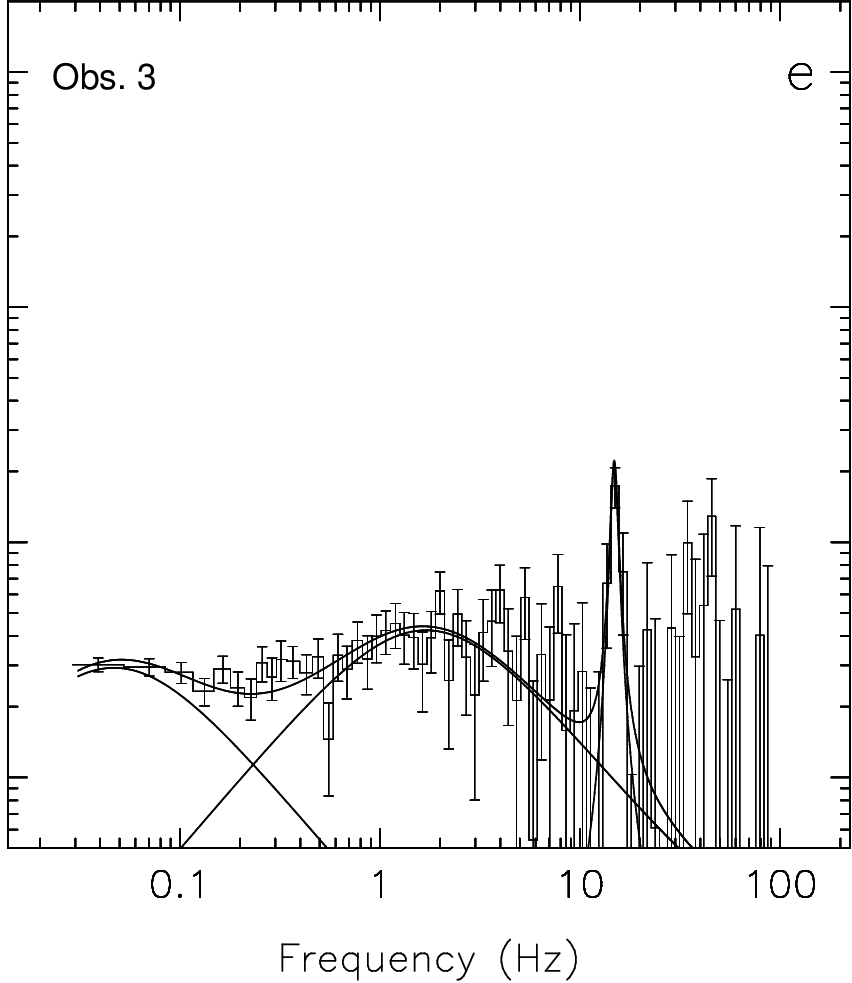}
\includegraphics[height=5.5cm]{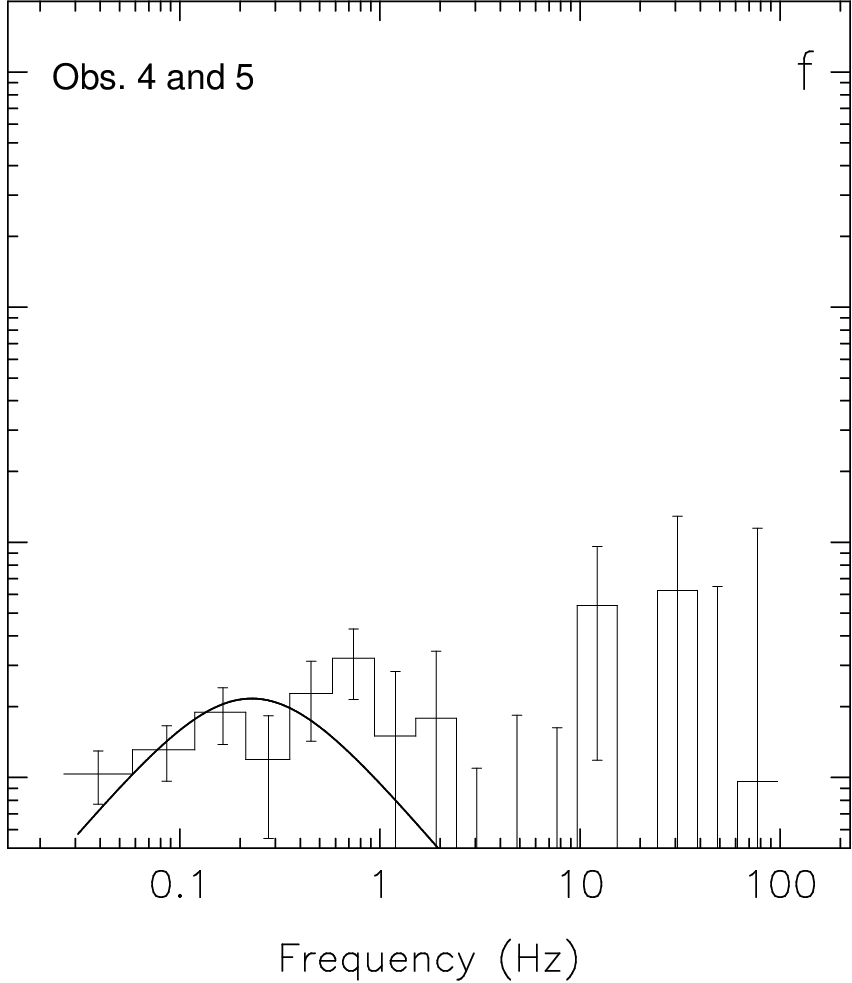}
\end{center}
\caption{Six examples of power spectra observed from H1743--322 in the 5.8--20.9 keV range. For all power spectra the Poisson noise was subtracted. The rebinning was varied between the panels to balance between the frequency resolution and the signal-to-noise ratio. The solid lines indicate the resulting fit function as well as the contribution of the individual Lorentzians.}\label{fig:pds}
\end{figure}

In the following we briefly summarize the power spectral properties
at frequencies below 100 Hz. The power spectra of our five
observations are shown in Figures \ref{fig:pds}b--f. For comparison
we have also included a representative power spectrum from the last
of the six observations made during the rise (Fig.~\ref{fig:pds}a). 

\noindent {\it Observations a--f} : in all six observations strong
band limited noise was observed, with a QPO around the break. The
frequency of this QPO increased from $\sim$0.06 Hz on March 28 to
$\sim$3.2 Hz on April 10. In Figure \ref{fig:pds}a we show the power
spectrum of the April 10 (MJD 52739) observation as a representative
example. The strength of the 0.016--100 Hz variability was
21.4$\pm$0.1\% rms, while that of the QPO at \numax=3.21$\pm$0.01 Hz
was 15.5$\pm$0.2\% rms. The QPO had a relatively high Q-value of
9.4$\pm$0.3. A sub-harmonic and second harmonic were present at
1.62$\pm$0.3 Hz  (Q$\sim$1.4) and 6.37$\pm$0.03 Hz (Q$\sim$9.8). The
measured phase lag between the photons in the 5.5--14.9 keV and
2--5.8 keV bands at the frequency of the 3.2 Hz QPO was
0.0019$\pm$0.0009 cycles (of 2$\pi$ radians), while the phase lags
for the  sub- and second harmonics were 0.013$\pm$0.006
and 0.012$\pm$0.003 cycles, respectively.
Positive values for the lags indicate that variations in the hard
flux are lagging those in the soft band. 

\noindent {\it Observation 1} :  the power spectrum showed red noise
with broad structures superposed (see Fig.~\ref{fig:pds}b). Three
zero-centered (i.e. Q=0) Lorentzians were used to fit the noise
continuum. Note that fitting the noise continuum with a broken
power-law  plus a Lorentzian did not result in an improvement. One
Lorentzian with \numax=0.015$\pm$0.007 Hz fits the red noise
component. The other two had \numax\ of 0.45$\pm$0.02 and 4.6$\pm$0.2
Hz. All three components had fractional rms amplitudes of around 5\%.
There were hints of a narrow QPO (Q$\sim$10) around 20 Hz, but it was
not significantly detected ($\sim$2$\sigma$). A power spectrum
with a lower frequency of $\sim5\times10^{-4}$ Hz revealed a peak in
the power spectrum  around 0.003 Hz when plotted in a $\nu P(\nu)$
representation (see Fig.~3 in \citet{miraho2004}), suggesting a
typical time scale of $\sim$300 s for the variations that are seen in Figure \ref{fig:flaring}.

\begin{figure}
\centerline{\includegraphics[height=12cm,angle=-90]{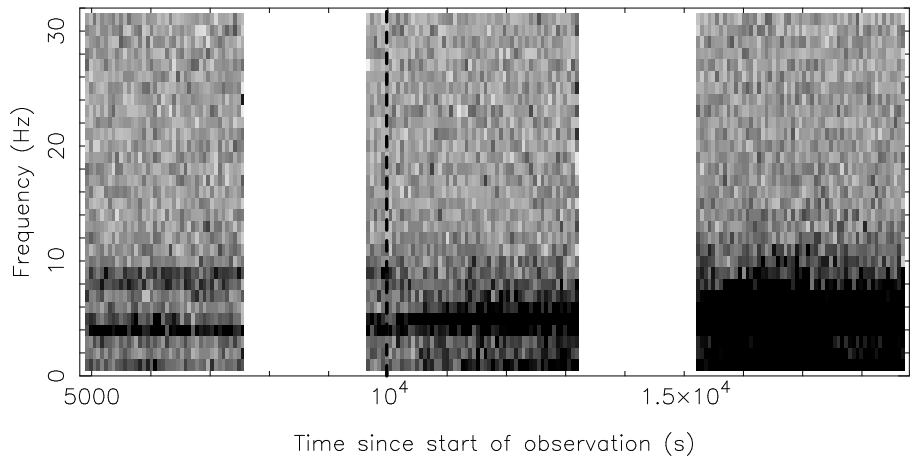}}

\caption{A dynamical power spectrum (5.8--20.9 keV) of orbits 2, 3
and 4 of observation 2. Time resolution is 64 seconds and frequency
resolution is 1 Hz. Darker grey indicates higher power. The two
white parts represent gaps in the data. The transition in color that can be seen in Figure \ref{fig:curves}b occurred around t=9980 s and is indicated in the figure by the dashed vertical line. Notice that the harmonic of the type-B QPO, which had a frequency of $\sim$9.3 Hz, disappeared shortly after this transition.}\label{fig:dyn}

\end{figure}

\begin{figure}
\vbox{
\centerline{\includegraphics[width=10cm]{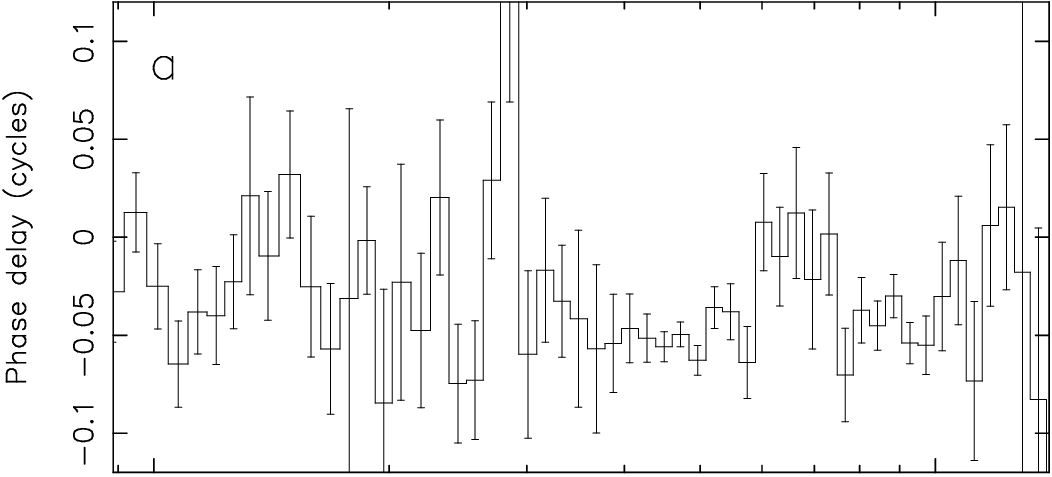}}
\centerline{\includegraphics[width=10cm]{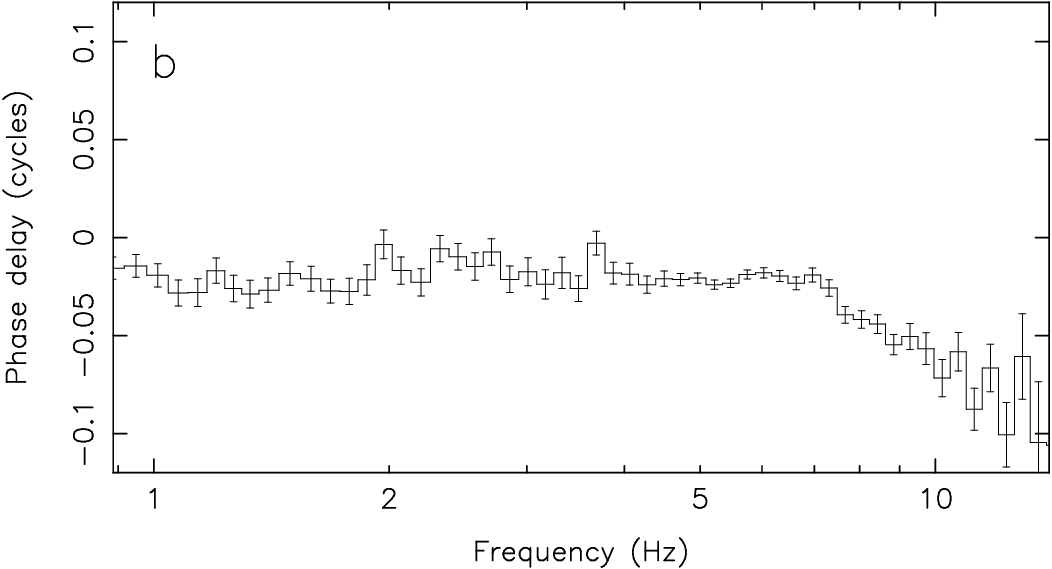}}
}
\caption{Phase lag spectra of the first (a) and second (b) part of observation 2. The phase lags are in units of cycles of 2$\pi$ radians and were calculated for the 2--5.8 keV and 5.8--14.9 keV bands. Negative values indicate that variations in the 2--5.8 keV band lag those in the 5.8--14.9 keV band.}\label{fig:phase}

\end{figure}

\noindent {\it Observation 2} : fast changes in variability
properties are known to happen in cases of a sudden change in
hardness and count rate, like the one seen in this observation (see
Fig.~\ref{fig:curves}). Inspection of the dynamical power spectrum
(part of which is shown in Figure \ref{fig:dyn}) revealed such
changes and we therefore decided to split the observation in two
parts, using the time of the sharp transition in color as the
divider. The power spectrum of the first part of the observation is
shown in Figure \ref{fig:pds}c. It is dominated by red noise and two
relatively sharp peaks. The noise was fitted with two zero-centered 
Lorentzians with \numax\ of 0.033$\pm$0.013 Hz and 0.44$\pm$0.07 Hz,
both with an rms of $\sim$4\%. The two QPOs at \numax=4.801$\pm$0.014
Hz and 9.27$\pm$0.04 had Q-values of $\sim$5.3 and $\sim$3.6,
respectively, and strengths of 5--6\% rms. Note that the two peaks
appeared not to be exactly harmonically related
(ratio=1.931$\pm$0.014), even not when using $\nu_0$ instead of
\numax. There were no large residuals in the 3--10 Hz range with our
fit function that might explain the ratio being different from 2. 
Nevertheless, we will refer to the second peak as the harmonic in the remainder of the paper. The two peaks had phase lags of
$-0.069\pm0.004$ cycles and $-0.055\pm0.005$ cycles (see also Fig.~\ref{fig:phase}a), indicating that
the soft photons were lagging the hard photons. Again, we found a
hint of a QPO around 20 Hz (\numax$\sim$22 Hz, rms$\sim$1.3\%, and Q
fixed to 10), which in this case was $\sim$3$\sigma$. 

The power spectrum of the second part (see Fig.~\ref{fig:pds}d) was
more complex. Especially the frequency range around the low frequency
QPOs was difficult to fit. The change in the power spectrum between
the first and the second part (see also Fig.~\ref{fig:dyn}) is best
described as follows: both the fundamental and its second harmonic
increased in frequency by about 15\%, to 5.53$\pm$0.02 Hz and
11.1$\pm$0.1 Hz. The fundamental became less coherent (Q$\sim$3.7)
and increased in strength to  $\sim$6.9\% rms, while the second
harmonic decreased in strength to $\sim$3.3\% rms with a similar Q
value. Notice that the harmonic was still visible in the dynamical
power spectrum, up to the point of the transition. In addition to
these changes a new component appeared at \numax$\approx$7.65 Hz,
which had an rms of $\sim$5\% and a Q value of $\sim$1.8. Also the
noise component needed the addition of an extra Lorentzian at
\numax$\approx$1.7 Hz (Q$\approx$0.8 and rms of $\sim$2.4\%), while
the other two noise components and the QPO at 22 Hz (now detected at
a 4.5$\sigma$, but a Q of 3.5)  remained relatively unchanged. The
phase lags of the 5.5 Hz and 11 Hz QPOs were  $-0.034\pm0.001$ and
$-0.069\pm0.002$ cycles, respectively. While the sign of the phase
lags of the two QPOs remained the same compared to the first part,
the overall phase lag spectrum seemed to change, as can be seen from
Figure \ref{fig:phase}b. Also, in the case of the QPOs in the second
part, no distinct features can be seen in the phase lag spectrum at
the frequencies of the QPOs, suggesting that the measured lags are
those of the noise continuum. Notice that the lags increased above
the frequency of the newly detected feature at 7.65 Hz.

\noindent {\it Observation 3} : the power spectrum (see
Fig.~\ref{fig:pds}e) was dominated by red noise, which was fitted
with two zero-centered Lorentzians (\numax=0.046$\pm$0.007 Hz, and
\numax=1.7$\pm$0.2 Hz) both with a strength of $\sim$1\% rms. A
narrow (Q=11$^{+7}_{-3}$) QPO was detected (4.2$\sigma$) at \numax=
14.84$\pm$0.18 Hz, with an rms of 1.7$\pm$0.2\%.

\noindent {\it Observations 4 \& 5} : the combined power spectrum of these
observations (see Fig.~\ref{fig:pds}f) was also dominated by red
noise. It was fitted with one zero-centered Lorentzian, which had a
\numax\ of 0.22$\pm$0.08 Hz and an rms of 2.6$\pm$0.2\%.

All the above fits had a reduced $\chi^2$ of less than 1.2, except
for the fits to the power spectra of observation f and the second
part of observation 2, which had a reduced  $\chi^2$ of 1.36 and
1.55, respectively. Analysis of other energy bands showed that all the QPOs became stronger toward higher energies.

\subsubsection{High frequency QPOs}

\begin{figure}
\begin{center}
\includegraphics[width=12cm]{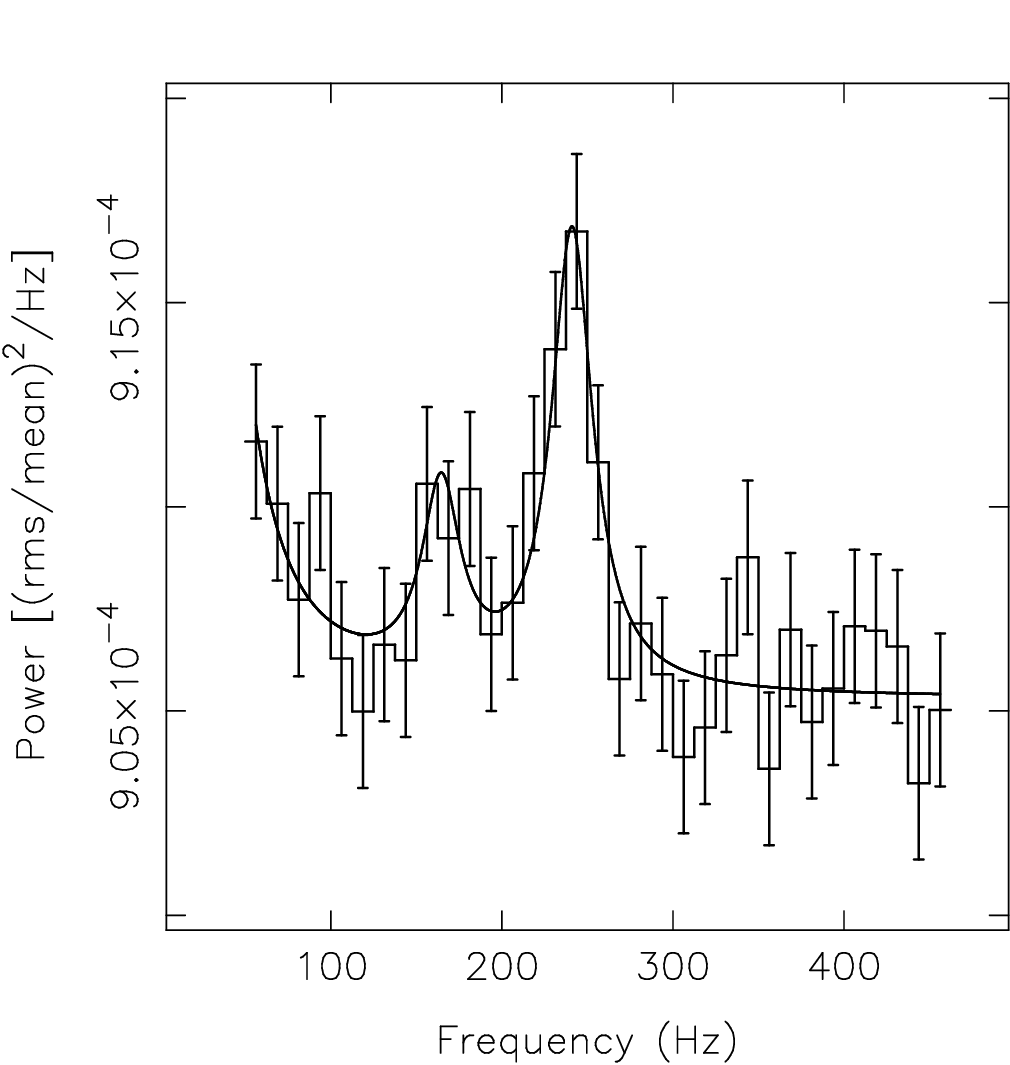}
\end{center}
\caption{The high frequency power spectrum (5.8--20.9 keV) of the second part of observation 2, showing QPOs at 165 and 241 Hz. The Poisson level was not subtracted. The solid line shows the best fits to the power spectrum with a power law and two Lorentzians. }\label{fig:hfq}
\end{figure}

All observations were searched for high frequency QPOs in the
5.8--20.9 keV band. As they are usually only found in observations
that show low frequency QPOs in the 5--10 Hz range we first searched
in observation 2. In that observation a high frequency QPO was
detected at \numax=240$\pm$3 Hz. It had an rms of
2.28$^{+0.34}_{-0.25}$\% (4.5$\sigma$ detection) and a Q-value of 7$\pm$2. Like in the case of the low
frequency part of the power spectrum, we found differences between
the first and second part of the observation. In the first part the
QPO is detected at 244$^{+26}_{-23}$ Hz, but only at a 1.8$\sigma$
level, with a Q of 2.6$^{+2.8}_{-1.4}$ and strength of
3.4$^{+2.1}_{-0.9}$\% rms. A fit to the second part (see
Fig.~\ref{fig:hfq}) gives \numax=241$\pm$3, Q=8$\pm$2, and an rms of
2.20$^{+0.28}_{-0.23}$\% (4.8$\sigma$ detection). Excess power at
lower frequencies ($\sim$150-180 Hz) was visible in this selection.
Fitting this with a second Lorentzian gave a very sharp (Q$\sim$17
peak at 160 Hz, which only seemed to fit the lower part of the
excess. We decided to fix the Q value to 6, forcing the Lorentzian to
fit the whole excess. This gave a frequency of 165$^{+9}_{-5}$ Hz and
an rms of 1.40$\pm$0.24 (a 3$\sigma$ detection). The two measured
frequencies had a ratio of 1.46$\pm$0.08, consistent with the 3:2
ratio that has been seen in several other black hole X-ray binaries.
We also measured the fractional rms amplitudes of the two QPOs in the
3.3--5.8 keV and 21.0-60 keV bands but only found upper limits:
$<$1.2\% and  $<$13\% for the 165 Hz QPO, and $<0.6$\% and $<$13\%
for the 241 Hz QPO. Additional selections (only orbits 5--7)
revealed  an excess below 100 Hz. We added a third Lorentzian to our
model but, while the obtained \numax of 78$\pm$5 Hz (Q$\sim$4) is
suggestive of frequency ratios of 1:2:3, the fit showed that the peak
itself was not significant ($<$2$\sigma$).

None of the other observations showed indications for high frequency
QPOs, but only for observation 1 were the upper limits on the
strength ($<1.8$\% rms) below the values measured in observation 2.

\section{Discussion}

We studied the variability properties of the black hole transient
\1746 in five RXTE observations that were performed simultaneously
with {\it Chandra} observations, as well as of six observations that
were taken at the start of the outburst. The source showed a wide
variety of variability properties, including high and low frequency
QPOs and variability on a time scale of a few hundred seconds. These
phenomena are discussed later in this section. First, by comparing
the variability properties with the spectral properties of the
observations we attempt to assign X-ray states. 

\subsection{X-ray states}

Observations a--f, at the start of the outburst (before the dashed
line in Fig.~\ref{fig:asm}), show a clear spectral evolution.
Preliminary spectral fits to the PCU-2 data \citep[see][for a
description of data reduction]{miraho2004} show that the spectra were
dominated by a power law component with an index increasing from
$\sim$1.4 in the first to $\sim$2.4 in the last of the six
observations. This, coupled with presence of the strong band limited
noise and QPOs with increasing frequencies of $\sim$0.06--$3.2$ Hz,
suggest that the source was in a transition away from the hard state
but had not yet reached the steep power law state. 

The broad band spectral properties of the five RXTE observations that
were performed (almost) simultaneously with {\it Chandra} are
tabulated in \citet{miraho2004}. All spectra were fitted with a
combination of a power law and a multi-color disk black body. The
fractional contribution of the power law to the 3--100 keV flux was
about 22\% for the observations 1 and 3, 50\% for the first part of
observation 2, $\sim$57\% for the second part, and  6\% for
observations 4 and 5 combined. Observations 4 and 5 are easily
classified as soft state (or thermal dominant state), given their low
power law flux and low level of variability. Observation 2 is in the
steep power law (or very high) state: the power spectrum shows
$\sim$6 Hz QPOs on top of moderate red noise while the energy
spectrum shows has a steep power law (index of 2.6) that contributes
about half of the total flux. We note however that, although it is
sometimes observed in the hard state, strong variability on a time
scale of a few hundred seconds is not typical of the steep power
law state (\citet{miraho2004} suggest that it may be related to the
high source inclination, see below). The same is true for observation 3, which
was likely in the soft state. Although the power law contribution is on the high side for the soft state and the power law index is
quite low ($\sim$2.1), the shape and strength of the broad band
variability and the presence of a weak QPO around 15 Hz are
consistent with the properties of the high luminosity soft state in
XTE J1550--564 \citep[see][]{howiva2001}. Observation 1 is hard to
classify. Spectrally it is similar to observation 3, but the broad
band noise is stronger. We note that the higher strength
is not due to variations  seen in Figure \ref{fig:flaring}, as that
occurred on a much longer time scale. Moreover, similar variability
was also seen in observation 3. Based on the spectral and variability
we suspect that this observation is intermediate between the steep
power law and soft states, although the origin of the broad noise
component around 0.5--5 Hz remains unclear.

\begin{figure}
\begin{center}
\includegraphics[width=12cm]{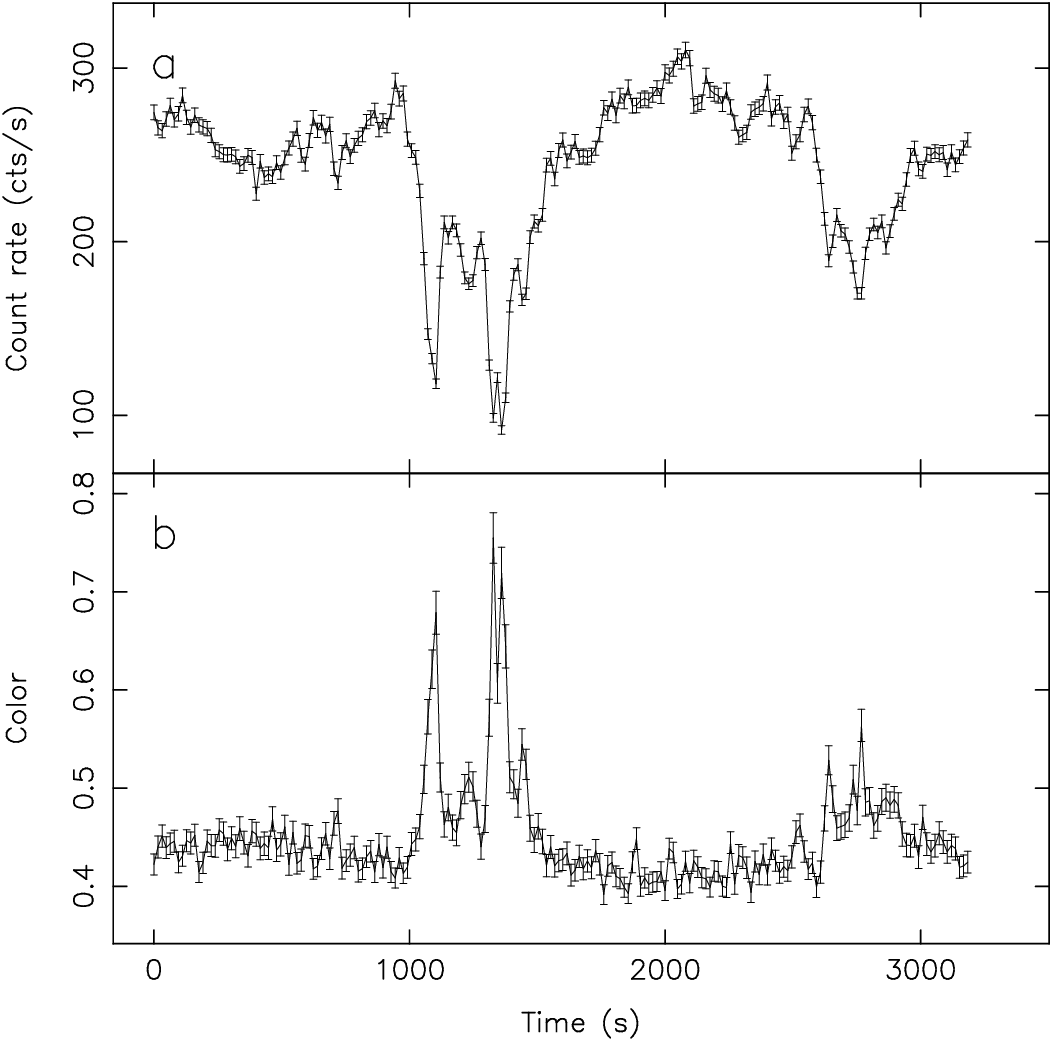}
\end{center}
\caption{(a) A 2--5 keV light curve and (b) color curve of observation 80146-01-40-00 (MJD 52776.6) of \1746\ showing strong dipping on a time scale of tens of seconds. Both curves are from background subtracted data and have time resolutions of 16 s. For a definition of color see Section \ref{sec:obs}.}\label{fig:dips}
\end{figure}

The above suggests that \1746 went through most of the known black
hole states. However, we did have some difficulty assigning states to
some of our observations. Inspection of data that have become public
while this work was in progress revealed strong hard dips in the
light curves, with up to 70\% decreases at low energies (see
Fig.~\ref{fig:dips}). These are similar to dips that have been
observed in GRO J1655--40 and 4U 1630--47 \citep{kuwibe1998} and
suggest a high source inclination ($\sim$60--70$^\circ$). Although a
more thorough analysis of those observations is beyond the scope of
this paper, a high inclination may explain why some of our
observations were difficult to classify.

%The power law contribution is relatively low, though not as low as
%expected for the soft state. The shape of the noise continuum is
%quite similar to that of the soft state observations of XTE
%J1550--564 during its 1999 high luminosity soft state
%\citep[see][]{howiva2001}, but its strength is a magnitude higher. We
%note that the higher strength is not due to flaring as seen in
%Figure\ref{fig:curves}, as that occured on a much longer time scale,
%although we cannot rule out that they are related. The (possible)
%presence of a weak QPO around 15--20 Hz also suggest a similarity to
%the soft state; QPOs like these have only been reported for XTE
%J1550--564 and GRO J1655--40 in their soft states. Given the shape of
%the continuum, the Obs 3 is soft state

\subsection{Low frequency variability}

Various types of low-frequency QPOs were observed, which we try to
identify with the three  types introduced by \citet{wihova1999} and
\citet{resomu2002}. Most of the observed QPO properties, such as
frequency, Q-values, rms and relation to noise shape/strength, are
similar to those observed in other black hole X-ray binaries. Note,
however, that the original type-A, B, C definition is solely based on
observations of XTE J1550--564 and detailed properties of these QPO
types may therefore by different in other sources. 

The strong 0.06--3.2 Hz QPOs on top of the band-limited noise in observations a--f were of type C. They showed a large range
in frequency, large amplitudes and Q-values, and both the fundamental
and second harmonic had hard lags. Only the phase lag of the
sub-harmonic had the opposite sign of that observed in XTE
J1550--564.

Observations 1--3 showed indications for weak QPOs between 14 and 22
Hz. Such QPOs have so far only been reported for XTE J1550--564 and
GRO J1655--40 in their soft states \citep{howiva2001,somcre2000a}. It
is not clear if these QPOs are related to any of the three more
commonly observed types. Most likely it is not a type-A or B QPO,
because it was detected simultaneously with QPOs of that type in
observation 2 and did not change its frequency when the other low
frequency QPOs did.  Fitting the noise continuum in the power
spectrum of observation 3 with a broken power law instead of two
Lorentzians gives a break frequency of $\sim$5 Hz. Using this
frequency and that of the QPO we find that the source falls on top of
the relation between break frequency and QPO frequency that was
found by \citet{wiva1999} for the black hole X-ray binaries and the Z
and atoll type neutron star sources. Note that the points for the
soft state QPO in  XTE J1550--564 lie close to the relation as
well. For that relation \citet{wiva1999} only used type-C QPOs from
black holes, suggesting that the weak 15--22 Hz QPOs in the much
softer states might be related to that type. The high Q-value of the
15 Hz QPO in observation 3 seems to support such an identification.

Observation 2 showed QPOs around 5--6 Hz which evolved during (and
after) a transition in count rate and color. Before the transition,
most of the QPO properties (frequency, strength and coherence) seem
to indicate type B. The only difference is the sign of the phase lag
of the fundamental, which was soft, whereas in XTE J1550--564 and XTE
J1859+226 it was found to be hard or consistent with zero. After the
transition the QPO properties indicate type A (type A-I to be more
specific - see \citet{howiva2001}). This suggests that for the first
time we have observed a smooth transition from a type-B to a type-A
QPO. It also indicates that the two types are more intimately related
then previously thought, possibly reducing the number of basic
mechanisms for low-frequency QPOs in the 1--10 Hz range to two.

Fast transitions in the light curves of black hole transients
(sometimes referred to as 'dips' and 'flip-flops') have been observed
in several other systems: GX 339--4 \citep{mikiki1991,nebeho2003}, GS
1124--68 \citep{tadomi1997}, XTE J1550--564 \citep{howiva2001}, 4U
1543--47 \citep{pamimc2004} and XTE J1859+226 \citep{cabeho2004}. In
all the cited cases the type of low frequency variability changed,
but, because of the shorter transition times, the evolution of the
QPO across the transition could not be followed. \citet{cabeho2004}
found that all of the transitions in XTE J1859+226 involved type-B
QPOs, which is also true in our case and in the cases reported for
XTE J1550--564, GX 339--4 \citep[in the transition reported
by][]{nebeho2003}, 4U 1543--47, and judging from the shape of the
QPOs, also in GS 1124--68. In XTE J1859+226 transitions from type-B
to type-A were always found to correspond to increases in the count
rate, while transitions from type-C to type-B correspond to decreases
in the count rate. Although an increase in count rate is also
observed in our case, the color evolution is opposite to that of the
B$\rightarrow$A transitions in XTE J1859+226 and the overall relation
found in XTE J1550--564, suggesting a more complex relation between
QPO type and spectral hardness. It is not clear whether differences
like these and different signs of the phase lags of the low frequency
QPOs, are related to a higher source inclination.

Three of the observations showed strong variations on a time scale of
a few hundred seconds. Although these variations are also seen in the
{\it Chandra} light curves, they become about twice as strong toward
higher energies. In observation 2 the $\sim$300 s variability seemed
to be stronger after the transition, which was also seen during a
similar transition in GX 339--4 \citep[see Figure 2
in][]{nebeho2003}. \citet{miraho2004} found that absorption lines in
the {\it Chandra} spectra were modulated by these variations, which,
combined with the suggested high inclination of the source, could
imply that the variations might be caused by absorbing or obscuring
structures in the observed disk outflow. In that case one would
naively expect the modulations to be stronger at lower energies or
energy independent, which is the opposite of what we observed.
However, variations in the disk outflow itself, which is highly
ionized, might lead to stronger variations at higher energies if the
wind is able to up-scatter photons to higher energies.  In
observation 3 the 15 Hz QPO is most significantly detected at the
lower count rates, which also suggests a more complex relation
between these QPOs and the $\sim$100 s variations than a purely a
line-of-sight effect. In fact, the latter suggests that the
variations might be related to the low frequency QPOs around
0.01--0.1 Hz, as seen in GRS 1915+105 \citep{moregr1997} and 4U
1630--47 \citep{dibe2000}. These are often simultaneously present
with QPOs at higher frequencies (1--10 Hz), whose frequencies seem to
be modulated by the 0.01--0.1 Hz QPOs. However, the dynamical power
spectrum of observation 2 (which also showed $\sim$100 variations)
only shows changes in the QPO frequency related to the transition and not to
the few hundred second variations.

\subsection{High frequency QPOs}

High frequency QPOs have recently received considerable attention
because of their apparent frequency distribution. In several sources
(GRO J1655--40, GRS 1915+105, XTE J1550--564, and possibly XTE
J1650--500) evidence has been found for  clustering of the high
frequency peaks  around frequencies which have ratios that are
consistent with 3:2, 3:2:1, or 5:3
\citep{st2001a,st2001b,miwiho2001,remumc2002a,remumc2002b,hoklro2003}.
Our detection of  simultaneous QPOs at $\sim$240 Hz and $\sim$160 Hz
in \1746\ strengthens the idea that a 3:2 ratio is a common feature
of high frequency QPOs in black hole X-ray binaries. We note that the
240 Hz and 160 Hz QPOs have also been detected  (at a higher
significance) in other observations of \1746\ by Remillard et al.
(2004, in prep.). In their case the QPOs were found by combining 26
and 9 observations, respectively, after selecting on $>$7 keV count
rate, but they were not detected simultaneously (Remillard, priv.\
comm.). This suggests shows that, like in XTE J1550--564 and GRO J16550--40 \citep{remumc2002a}, the relative strength of the two harmonically related QPOs \1746\ is variable, to the extent that at times only one of them is visible. 

Several models have been proposed to explain the frequencies of these
high frequency features, most of them assuming a disk origin of the
oscillations
\citep{nowabe1997,stvimo1999,cuzhch1998,psno2001,lati2001,abkl2003,reyoma2003,scbe2004}.
Only in the last three of these references possible explanations of  
the observed frequency ratios are offered. \citet{abkl2003} discuss
two possible models for the observed 3:2 ratio (they do not discuss
the 5:3 ratio observed in GRS 1915+105): the first is the presence of
forced 3:1 and 2:1 vertical and radial resonances through pressure
coupling (which requires the additional presence of combination
frequencies to explain the observed 3:2 ratio) and the second one is
parametric resonance between random fluctuations in the (epicyclic)
vertical and radial directions. \citet{reyoma2003} explain the QPOs
as trapped pressure mode oscillations in a torus orbiting a black
hole, whose  eigenfrequencies are found to being 2:3:4 harmonics.
These three resonance models, and in fact most other models,  do not
explain how the X-ray light curves become modulated at the calculated
frequencies. An exception is the model by \citet{scbe2004}. These
authors used a ray tracing code in a Kerr metric to explore the light
curves of isotropically emitting massive test particles orbiting a
rotating black hole. Reasonable amplitudes are retrieved for the
simulated modulations, with higher amplitudes being expected for
higher viewing angles. The latter seems to be confirmed by the
detection of high frequency QPOs in GRS 1915+105, GRO J1655--40, XTE
J1550--564, XTE J1859+226, 4U 1630--47, XTE J1650--500 and \1746, all
sources which are believed to have moderate to high inclinations
($\sim$50--75$^\circ$). Assuming a preference for the particles to
occur at or near closed orbits (where $\nu_r$, $\nu_\phi$, and
$\nu_\theta$ are rational multiples of each other) the authors are
able (not surprisingly) to create modulations at 1:2:3 ratio. More
importantly, however, is the elegant way in which they produce
changes in the relative strength of these harmonically related peaks
\citep[see][and this work]{remumc2002a}, by varying the arc-length of
the emission region. In the models of \citet{abkl2003} and
\citet{scbe2004} the QPO frequencies are determined by the spin and
mass of the black hole. Estimates of the spin can be made when a mass
function of the black hole is available, which is unfortunately not
the case yet for \1746. 

The transition seen in observation 2 provides opportunity to test the
relation between the low and high frequency QPOs, since the low
frequency QPOs showed significant changes in their frequency. During
the transition the frequency of the low frequency QPOs increased by
$\sim$15\%. Unfortunately, the high frequency QPO before the
transition was not detected at a high significance. However, its
frequency (244$^{+26}_{-23}$ Hz) is consistent with the value
obtained after the transition (241$\pm$3 Hz) and only marginally
consistent with 205 Hz (which is 241 Hz -- 15\%). If the observed
constancy of the high frequency QPO is real, the frequency increase
of the low frequency QPO suggests that the latter is not due to
Lense-Thirring precession, as suggested by \citet{scbe2004}, since
in that case the higher frequency QPO should have changed in frequency  as well. 

\acknowledgements

JH wishes to thank Ron Remillard and Jeff McClintock for sharing
their results prior to publication and Andy Fabian for comments on an
early version of the paper. JH and WHGL gratefully acknowledge
support from NASA. JMM gratefully acknowledges support from the NSF
through its Astronomy and Astrophysics Postdoctoral Fellowship
program. TB was partially supported by MUIR under CO-FIN grant
2002027145. DS acknowledges support from the SAO Clay Fellowship. 
This research has made use of the data and resources obtained through
the HEASARC on-line service, provided by NASA-GSFC.

\newpage
\clearpage

%\bibliographystyle{aa}
%\bibliography{all-bib}

\begin{table}
\caption{Log of our five RXTE/PCA observations of \1746}\label{tab:log}
\begin{center}
\begin{tabular}{cccccc}
\hline
\hline
Obs. & Date & Start-Stop$^a$  & Total & counts $s^{-1}$ & rms$^b$ \\
     &      & (UTC)           & (ks)  &       & (\%) \\
\hline
1 & 2003-05-01 & 17:00--02:03 & 24.3 & 1760.0$\pm$0.2 & 10.19$\pm$0.10 \\
2 & 2003-05-28 & 05:29--16:53 & 24.6 & 2037.4$\pm$0.3 & 10.87$\pm$0.09\\
  &            & before       & 4.9  & 1824.1$\pm$0.6 & 10.02$\pm$0.15\\
  &            & after        & 19.7 & 2091.2$\pm$0.3 & 11.11$\pm$0.09\\
3 & 2003-06-23 & 17:05--22:43 & 13.7 & 1120.4$\pm$0.3 & 5.4$\pm$0.3\\
4 & 2003-07-29 & 20:54-22:51  & 4.7  & 752.1$\pm$0.6 & 5.3$\pm$0.8$^c$ \\
5 & 2003-07-30 & 19:49--00:43 & 2.4  & 757.1$\pm$0.7 & 5.3$\pm$0.8$^c$\\
\hline
\end{tabular}
\end{center}
\noindent $^a$ Observations 1 and 5 ended the next day. Before and after for observation 2 refers to before and after the transition in the light curve.\\ $^b$ Fractional rms amplitude (5.8--20.9 keV) in the 0.016--100 Hz range \\ $^c$ Measured from the combined power spectra of observations 4 and 5 \\

\end{table}

\newpage
\clearpage

\end{document}